\documentclass{article} 
\usepackage{mlgenx_conference,times}
\usepackage{amssymb,amsmath,amsthm}

\newtheorem{theorem}{Theorem}

\usepackage{mathtools}

\usepackage{amsmath,amsfonts,bm}

\def\eqref#1{equation~\ref{#1}}

\def\1{\bm{1}}

\DeclareMathAlphabet{\mathsfit}{\encodingdefault}{\sfdefault}{m}{sl}
\SetMathAlphabet{\mathsfit}{bold}{\encodingdefault}{\sfdefault}{bx}{n}

\usepackage{hyperref}
\usepackage{url}
\usepackage{tikz}
\newcommand{\indep}{\perp \!\!\! \perp}
\usepackage{algorithm,algpseudocode}
\usepackage{subcaption}
\usepackage{comment}

\title{Multi-omic Causal Discovery using Genotypes and Gene Expression}

\author{Stephen Asiedu \\ \textbf{David Watson}  \\
Department of Informatics\\
King's College London\\
\texttt{\{stephen.asiedu, david.watson\}@kcl.ac.uk} \\
}

\newtheorem{definition}{Definition}

\iclrfinalcopy 
\begin{document}

\maketitle

\begin{abstract}
    Causal discovery in multi-omic datasets is crucial for understanding the bigger picture of gene regulatory mechanisms but remains challenging due to high dimensionality, differentiation of direct from indirect relationships, and hidden confounders. We introduce GENESIS (GEne Network inference
    from Expression SIgnals and SNPs), a constraint-based algorithm that leverages the natural causal precedence of genotypes to infer ancestral relationships in transcriptomic data. Unlike traditional causal discovery methods that start with a fully connected graph, GENESIS initializes an empty ancestrality matrix and iteratively populates it with direct, indirect or non-causal relationships using a series of provably sound marginal and conditional independence tests. By integrating genotypes as fixed causal anchors, GENESIS provides a principled ``head start'' to classical causal discovery algorithms, restricting the search space to biologically plausible edges. We test GENESIS on synthetic and real-world genomic datasets. This framework offers a powerful avenue for uncovering causal pathways in complex traits, with promising applications to functional genomics, drug discovery, and precision medicine.
\end{abstract}

\section{Introduction}
In recent years, high-throughput technologies have generated vast volumes of multi-omic data across different layers such as transcriptomics, proteomics, and metabolomics \citep{qiao2024recent, reuter2015high, Manel2016Genomic}. While this wealth of information holds immense potential for advancing our understanding of biological systems \citep{Abu-Elmagd2022Editorial:, bourne2015perspective}, it simultaneously poses formidable analytical challenges, particularly in the realm of complexity and high-dimensionality \citep{Hu2018ApplicationOC}. Although considerable progress has been made in applying causal discovery methods within individual omics layers, these approaches fail to capture the intricate interplay between the different molecular networks. Integrating information across multiple omics layers promises to reveal more detailed explanations that would remain obscured if we solely rely on single-omics analyses \citep{Danchin2007The, Veenstra2012Metabolomics:, Neale2019Gene}. Adopting a multi-omic causal discovery approach is essential to advance our biological understanding and design more targeted therapeutic interventions \citep{He2017Big, mohammadi2023omics}. 

A major challenge in modern genomics is to infer the gene regulatory networks (GRNs) that dictate cellular behavior \citep{karlebach2008modelling}. A clear and precise understanding of GRNs can illuminate the pathways that lead to complex traits and diseases. However, the underlying data is inherently high-dimensional, posing major statistical and computational challenges. One promising strategy, which we build on below, is the usage of cis-expression quantitative trait loci (cis-eQTLs) \citep{michaelson2009detection}, which leverage two key biological facts: (1) that genetic variation precedes transcriptomic variation; and (2) that the influence of a genetic variant on a target gene decreases as a function of spatial proximity. 
This targeted approach boosts statistical power by exploiting prior knowledge and reducing the search space.

In this context, we introduce GEne NEtwork inference from expression SIgnals and Single nucleotide polymorphisms (GENESIS), an algorithm to integrate genotype and transcriptomic data to reconstruct directed GRNs. Our method harnesses the natural variability of SNPs to distinguish between direct and indirect gene regulatory effects by following the inference rules proposed by \cite{magliacane2016ancestral}. We implement a two-step hypothesis testing framework to identify marginal SNP-gene associations within cis-windows and filter out indirect effects via conditional independence tests. 
We use parametric plug-ins based on linear modeling assumptions that are widely used in bioinformatics \citep{Smyth_2004, love2014moderated}, ensuring the efficiency of our method. 
GENESIS is provably sound, delivering a partially oriented ancestrality matrix in polynomial time that can lead to major speedups when used as a preprocessing step for classical causal discovery methods like the PC algorithm  \citep{spirtes2001causation}. We illustrate our method on simulated and real-world data, where it compares favorably to the state of the art.


\section{Background}
We use upper case italics $X$ to denote random variables or sets thereof. 
Let $\mathcal{G} = \langle \mathcal V, \mathcal E \rangle$ be a graph with nodes $\mathcal V$ that represent variables and directed edges $\mathcal{E}: \mathcal V \times \mathcal V$ that denote causal relationships between them. 
We focus in particular on directed acyclic graphs (DAGs), as is common in causal discovery \citep{spirtes2001causation, peters2017elements}.
In the context of this study, nodes may represent SNPs (background variables $Z$) or genes (foreground variables $X$). We aim to draw inferences about the relationships between the latter by exploiting signals from the former.
Specifically, our goal is to infer as much as possible about the subgraph $\mathcal G_X = \langle X, \mathcal E_X \rangle$, with edge set $\mathcal E_X: \{Z \cup X\} \times X$ including all directed arrows into foreground variables.

We use kinship terms to describe relationships between nodes, with $Pa(\cdot), Ch(\cdot), An(\cdot), De(\cdot)$ representing the parents, children, ancestors, and descendants (respectively) of a given node set. 
If $X \in An(Y)$ (or, equivalently, $Y \in De(X)$), we write $X \prec Y$ (equivalently, $Y \succ X$). 
If $X \not\in De(Y)$, we write $X \preceq Y$.
We write $X \sim Y$ when neither variable is an ancestor of the other. 
Ancestry graphs impose a strict partial order on nodes, characterized by the following properties: (1) \textit{irreflexive}: $X \prec X \Rightarrow \texttt{FALSE}$; (2) \textit{asymmetric}: $X \prec Y \Rightarrow Y \not\prec X$;  and (3) \textit{transitive}: $X \prec Y ~\&~ Y \prec Z \Rightarrow X \prec Z$. 

We use standard probabilistic definitions of independence, writing $X \indep Y \mid Z$ to indicate that variable sets $X$ and $Y$ have no mutual information after conditioning on the (potentially empty) variable set $Z$. We assume that distributions are Markov and faithful to the underlying graph $\mathcal G$, in which case conditional independence claims are equivalent to $d$-separation statements \citep{pearl2009causality}. 

Building on the work of \cite{claassen2012logical} and \cite{watson2022causal}, we introduce the concept of (de)activators.
\begin{definition}[Deactivator]
A variable \( W \) is a \textit{deactivator} of the relationship between \( X \) and \( Y \) given \( Z \) if (a) $X \not\indep Y \mid Z$; and (b) $X \indep Y \mid Z \cup W$.
In this case, we write \( X \indep Y \mid Z \cup [W] \).
\label{def:minimal_deactivator}
\end{definition}
A deactivator is a single variable that, when added to the conditioning set $Z$, is sufficient to block all otherwise open paths linking $X$ and $Y$. 

\begin{definition}[Activator]
A variable \( W \) is an \textit{activator} of the relationship between \( X \) and \( Y \) given \( Z \) if (a) $X \indep Y \mid Z$; and (b) $X \not\indep Y \mid Z \cup W$. In this case, we write \( X \not\indep Y \mid Z \cup [W] \).
\label{def:minimal_activator}
\end{definition}
An activator is a single variable that, when added to the conditioning set $Z$, is sufficient to open an otherwise blocked path between $X$ and $Y$.


\section{Method and Algorithm}
In this section, we present the oracle version of GENESIS, which is designed to infer causal relationships among a set of variables through a careful series of conditional independence (CI) queries. Unlike traditional methods that begin with a fully connected graph and progressively remove edges (e.g., the PC algorithm), \textsc{GENESIS‑Oracle} is initialized with an empty ancestrality matrix $\mathbf M$ and incrementally adds causal information based on the results of calls to the oracle. Inputs include a background set $Z$ (e.g., SNPs or other exogenous data) and a foreground set $X$ (e.g., gene expression data), while the output is the ancestrality matrix $\mathbf M$ in which each element $\mathbf M_{ij}$ encodes the causal relationship (if decidable) between pairs of foreground variables $X_i, X_j$.

GENESIS relies on three simple inference rules, variants of which have appeared in several causal discovery methods \citep{claassen2012logical, entner2013data, watson2022causal}.
Let $A$ and $\{X, Y\}$ be two sets of nodes in the graph $\mathcal G$ where $A \preceq \{X, Y\}$, and let $A_{\backslash W} := A \backslash \{W\}$ for some node $W \in A$. 
Our first rule detects causal pathways via deactivation patterns:
\begin{itemize}
    \item[(R1)] If $\exists W \in A: W \indep Y \mid A_{\backslash W} \cup [X]$, then $X \prec Y$.
\end{itemize}
This is only possible when $W$ is a mediator on the path from $X$ to $Y$.
Our second rule rejects causal pathways via activation patterns:
\begin{itemize}
    \item[(R2)] If $\exists W \in A: W \not\indep X \mid A_{\backslash W} \cup [Y]$, then $X \preceq Y$.
\end{itemize}
This is only possible when $Y$ is a (descendant of a) collider on the path from $W$ to $X$ and is not a non-collider on any other path active under $A_{\backslash W}$.
Finally, our third rule establishes causal independence via $d$-separation:
\begin{itemize}
    \item[(R3)] If $X \indep Y \mid A$, then $X \sim Y$.
\end{itemize}

See Alg. \ref{alg:genesis-oracle} for a summary of the oracle procedure. We use the grow-shrink algorithm to infer the Markov blanket of each $X$, as this method is efficient and sound \citep{margaritis1999bayesian}. Alternatives are possible in practice, e.g. IAMB \citep{tsamardinos2003algorithms}. Keeping a running tab of each variable's Markov blanket can lead to major speedups over alternative procedures such as the confounder blanket learner \citep{watson2022causal}, which must cycle through vast conditioning sets for each pair of foreground variables. 

Once Markov blankets are initialized, the basic procedure is to cycle through all variable pairs. By (R3), we conclude that any foreground variables that are $d$-separated by their combined Markov blanket $S$ must be causally unconnected. Next, we loop through the elements of $S$, applying (R2) and (R3) to test for patterns of (de)activation that can help orient causal relations within $X$. Once we have done this for all pairs, we use a set of closure rules (see Alg. \ref{alg:closure} in Appx. \ref{app:closure}) that exploits the strict partial order on $\mathcal G_X$ to potentially draw some extra inferences about $\mathbf M$. Finally, we update the Markov blanket for each foreground variable in case any newly inferred non-descendants might enter in. The algorithm converges either when the ancestral graph is fully oriented or a complete pass fails to draw any new inferences.

We establish some basic properties of this algorithm. (For proofs, see Appx. \ref{app:proofs}.)
\begin{theorem}[Soundness]\label{thm:soundness}
    All inferences returned by {\sc GENESIS-Oracle} hold in the true $\mathcal G_X$. Moreover, if $\mathbf M_{ij} = i \prec j$, then the set of combined Markov blankets $S = MB(X_i) \cup MB(X_j)$ is a valid adjustment set for $(X_i, X_j)$.
\end{theorem}
This result follows from the soundness of the inference rules themselves, which has been previously established by numerous authors. 
\begin{theorem}[Complexity]\label{thm:complexity}
    Let $d_Z, d_X$ be the dimensionality of the background variables $Z$ and foreground variables $X$, respectively. Then the complexity of {\sc GENESIS-Oracle} is $\mathcal O(d_Z d_X^2)$.
\end{theorem}
This result holds regardless of graph density. In sparse settings, runtime can be highly efficient due to the relatively low cardinality of $S$.

In summary, our method builds its causal structure from the ground up by gradually adding evidence-based edges. This forward-construction approach not only avoids the exhaustive edge assignment typical of fully connected initializations but also supplies a robust starting point for downstream causal discovery algorithms, such as FCI or the PC algorithm. By constraining further searches to only those edges consistent with the inferred ancestry matrix, our method significantly improves both speed and reliability, particularly in high-dimensional genomic settings where disentangling direct and indirect regulatory effects is critical.

\begin{algorithm}[ht]
\caption{\sc{GENESIS-Oracle}}
\label{alg:genesis-oracle}
\textbf{Input}: Background variables $Z$, foreground variables $X$ \\
\textbf{Output}: Ancestrality matrix $\mathbf M$ \\
\begin{algorithmic}

\State Initialize: \texttt{converged} $\gets$ \texttt{FALSE},  $\mathbf M \gets [\texttt{NA}]$
\ForAll{$X_i \in X$}
\State $MB(X_i) \gets \texttt{GrowShrink}(X_i, Z)$
\EndFor
\While{\textbf{not} \texttt{converged}} 
  \State \texttt{converged} $\gets$ \texttt{TRUE}
  \ForAll{$(X_i, X_j) \in X $ s.t. $i < j, \mathbf M_{ij} = \texttt{NA} $}
    \State $S \gets MB(X_i) \cup MB(X_j)$
    \If{$X_i \indep X_j \mid S$} 
      \State  $\mathbf M_{ij} \gets$ $i \sim j$, \texttt{converged} $\gets$ \texttt{FALSE} 
    \Else
        \For{ $W \in S$ }
            \If{$W \indep X_j \mid S_{\backslash W} \cup [X_i]$}
              \State  $\mathbf M_{ij} \gets$ $i \prec j$, \texttt{converged} $\gets$ \texttt{FALSE}
            \ElsIf{$W \indep X_i \mid S_{\backslash W} \cup [X_j]$}
              \State  $\mathbf M_{ij} \gets$ $j \prec i$, \texttt{converged} $\gets$ \texttt{FALSE}
            \ElsIf{$W \not\indep X_j \mid S_{\backslash W} \cup [X_i]$}
              \State  $\mathbf M_{ij} \gets \mathbf M_{ij} \land j \preceq i$, \texttt{converged} $\gets$ \texttt{FALSE}
            \ElsIf{$W \not\indep X_i \mid S_{\backslash W} \cup [X_j]$}
            \State  $\mathbf M_{ij} \gets \mathbf M_{ij} \land i \preceq j$, \texttt{converged} $\gets$ \texttt{FALSE}
            \EndIf 
        \EndFor
    \EndIf
  \EndFor
  \State $\mathbf M \gets \texttt{Closure}(\mathbf M)$
  \ForAll{$X_i \in X$}
  \State $A_i \gets MB(X_i) \cup \{ W : W \preceq_{\mathbf M} X_i\}$
\State $MB(X_i) \gets \texttt{GrowShrink}(X_i, A_i)$
\EndFor
\EndWhile

\end{algorithmic}
\end{algorithm}

\begin{figure}[h!] 
  \begin{subfigure}{0.48\textwidth} 
    \includegraphics[width=\textwidth]{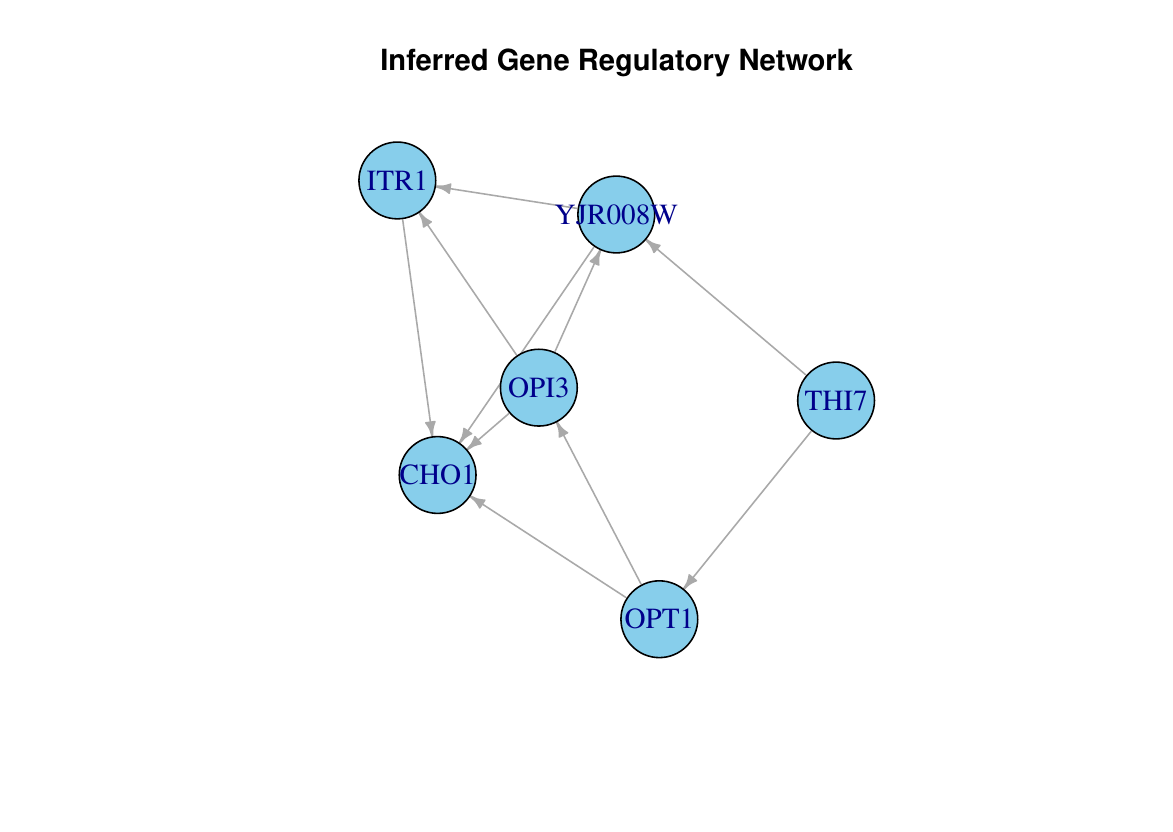} 
    \caption{Gene regulatory network for the phosphocoline subnetwork in \textit{Saccharomyces cerevisiae}  predicted by GENESIS}
    \label{fig:phosphocoline}
  \end{subfigure}
  \hfill 
  \begin{subfigure}{0.48\textwidth} 
    \includegraphics[width=\textwidth]{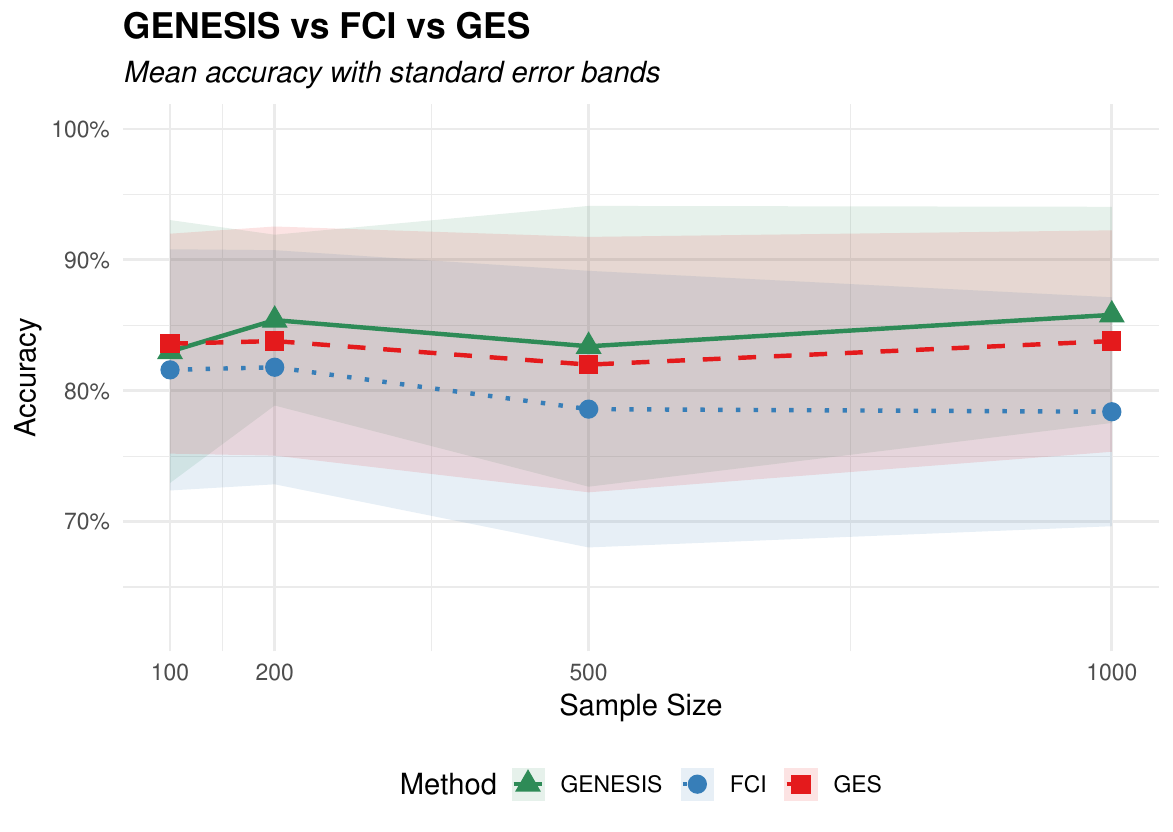} 
    \caption{A multivariate comparison of GENESIS with FCI (constraint-based algorithm) and GES (Score-based algorithm) }
    \label{fig:multivariate}
  \end{subfigure}
  \caption{Results on real world and simulated data}
  \label{fig:results}
\end{figure}

\section{Results and Conclusion}

We analyzed the yeast dataset provided by the TRIGGER package \citep{chen2007harnessing}, which comprises 112 recombinant haploid segregant strains derived from a cross between two haploid parental strains of Saccharomyces cerevisiae (BY and RM) \citep{brem2005landscape}. This dataset includes genome-wide expression profiles for 6,216 genes and genotypic information from 3,244 single-nucleotide polymorphism (SNP) markers. One key strategy we employed was the analysis of cis-expression quantitative trait loci (cis-eQTLs), which exploits the physical proximity between SNPs and their associated genes. By limiting our search to a 5 kilobase genomic window, which is a standard practice in cis-eQTL mapping, we prioritised genetic variants most likely to exert direct regulatory effects. Our investigation focused exclusively on genes previously identified in the literature as components of the phosphocholine subnetwork, which governs the biosynthesis of phosphatidylcholine  through the Kennedy pathway. This pathway is essential for maintaining membrane integrity and supporting critical cellular processes in yeast \citep{henneberry2001phosphatidylcholine}. To assist in identifying our iterative proximal ancestor sets, we utilized partial correlation tests within the GENESIS framework. The resulting regulatory network, shown in Figure~\ref{fig:phosphocoline}, adheres to an acyclicity constraint and captures ancestral relationships that are consistent with those reported in previous work on the yeast phosphocholine network \citep{chen2019inferring}.

To evaluate the performance of GENESIS in a controlled multivariate setting, we conducted a series of simulations using randomly generated directed acyclic graphs (DAGs) with both background $Z$ and foreground $X$ variables with the goal of inferring ancestral relationships $X$. The DAGs were randomly generated with varying sample size ranging between 100 and 1000 with edge densities chosen to maintain moderate sparsity. Our data were simulated from these structures using a linear Gaussian model with mixed noise. For each sample size, we compared the structural recovery accuracy of GENESIS against two well-established causal discovery methods: the Fast Causal Inference (FCI) \citep{spirtes2001causation} algorithm and Greedy Equivalence Search (GES) \citep{chickering2002optimal}. GENESIS was run with a conditional independence threshold of $\alpha = 0.05$ while FCI and GES were configured using default parameters. Performance was quantified as the mean adjacency matrix accuracy over 20 replicates, computed as the element-wise match between the inferred and true adjacency matrices among target variables. Across the sample sizes, GENESIS consistently performed better than the benchmark methods.

While GENESIS demonstrates strong empirical performance in both simulated and biological settings, its reliance on iterative heuristics and sensitivity to initialization highlight the need for further theoretical analysis and optimization. Future work may explore extensions to nonlinear models and formal guarantees for convergence, completeness, and identifiability under broader conditions.

\newpage
\bibliography{mlgenx_conference}
\bibliographystyle{mlgenx_conference}
\appendix

\section{Closure}\label{app:closure}
Below is the pseudocode for the closure described in Algorithm \ref{alg:genesis-oracle}.
\setcounter{algorithm}{1}
\begin{algorithm}[h]
    \small
   \caption{{\sc Closure}}
   \label{alg:closure}
\begin{algorithmic}
   \State {\bfseries Input:} Ancestrality matrix $\mathbf{M}$
   \State {\bfseries Output:} Updated ancestrality matrix $\mathbf{M}$
   \State
   \For{$i, j \in \{1, \dots, d_X\} ~\text{such that}~ i > j$}
   \If{$i \preceq_{\bf M} j \land i \succeq_{\bf M} j \lor i \sim_{\bf M} j$}
            \State $\mathbf{M}_{ij} \gets i \sim j$
    \ElsIf{$i \prec_{\bf M} j$}
            \State $\mathbf{M}_{ij} \gets i \prec j$
     \ElsIf{$j \prec_{\bf M} i$}
            \State $\mathbf{M}_{ij} \gets j \prec i$
    \EndIf
    \EndFor
    \State $\texttt{converged} \gets \texttt{FALSE}$
   \While{\textbf{not} $\texttt{converged}$}
   \State $\texttt{converged} \gets \texttt{TRUE}$
   \For{$i, j, k \in \{1, \dots, d_X\} ~\text{such that}~ i \neq j \neq k, i > k$}     
   \If{$i \prec_{\bf M} j \prec_{\bf M} k \land \mathbf{M}_{ik} \neq i \prec k$}
       \State $\mathbf{M}_{ik} \gets i \prec k, \texttt{converged} \gets \texttt{FALSE}$
    \ElsIf{$k \prec_{\bf M} j \prec_{\bf M} i \land \mathbf{M}_{ik} \neq k \prec i$}
    \State $\mathbf{M}_{ik} \gets k \prec i, \texttt{converged} \gets \texttt{FALSE}$
     \EndIf
   \EndFor
   \EndWhile
\end{algorithmic}
\end{algorithm}

\section{Proofs}\label{app:proofs}

\paragraph{Theorem 1 (Soundness)} 
\begin{proof}

By construction, \textsc{GENESIS-Oracle} only applies the three sound rules (R1), (R2) and (R3) to the union of the Markov blankets of $X_i$ and $X_j$. However, we feed the 
\texttt{GrowShrink} algorithm with only known non-descendants and hence the guarantee that the union of Markov blankets will be non-descendants themselves. We then apply closure under transitivity and asymmetry as seen in Alg \ref{alg:closure}. 
This means the soundness of our oracle depends on the soundness of the rules themselves. (R1) and (R2) follows from a direct application of Lemma 1 from \cite{magliacane2016ancestral} while (R3) is the direct application of faithfulness since we are limited to non-descendants. \\

For us to arrive at $M_{ij}=i \prec j$, we must use (R1) with some $S = MB (X_i) \cup MB(X_j)$ to detect a minimal deactivation of the form $W\indep X_j \mid S_{\backslash W} \cup [X_i]$ for some $W \in S$ as proved by \cite{entner2013data} (where $S$ is limited to a set of non-descendants). As stated earlier, the union of Markov blankets $S$ of $(X_i, X_j)$ will solely contain non-descendants as we start with $Z$ which is biologically a well establish non-descendant of $X_i \in X$ and $W \in S$. Since this satisfies the assumption of \cite{entner2013data}, we conclude, using the same argument, that $S = MB (X_i) \cup MB(X_j)$ is a valid adjustment set for $(X_i, X_j)$.
\end{proof}

\paragraph{Theorem 2 (Complexity)}
\begin{proof}
    From the {\sc GENESIS-Oracle} \ref{alg:genesis-oracle}, the initialization of the Markov blanket will cost $\mathcal O (d_X d_Z)$ as each foreground variables requires a pass through the background variables.
    Each unordered pair $(X_i,X_j)$, with $i<j$, is resolved at most once per execution of the \texttt{while} loop. There are $\binom{d_X}{2}=\mathcal{O}(d_X^2)$ such pairs in total for the pairwise inference loop. Initially since $S$ will contain at most $d_Z$ elements, it will cost $\mathcal{O} (d_Z)$ and $\mathcal{O} (d_Z + d_X)$ in subsequent iterations. We assume that $d_Z \gg d_X$, in which case this reduces to $\mathcal O(d_Z)$. In the oracle setting, the CI queries execute in constant time, $\mathcal O(1)$, although practical implementations tend to scale with the size of the conditioning set. 
    Thus the pairwise inference loop will cost $\mathcal{O}(d_Z) \times \mathcal{O}(d_X^2) = \mathcal{O}(d_Zd^2_X)$. 
\end{proof}

\end{document}